\documentclass[preprint,pre,showkeys,showpacs]{revtex4}
\usepackage{graphicx}
\usepackage{graphics}
\usepackage{latexsym}
\usepackage{subfigure}
\usepackage{multirow}
\newcommand{\beq}{\begin{equation}}
\newcommand{\eeq}{\end{equation}}
\begin{document}

\title{The dynamical origin of the universality classes of spatiotemporal intermittency}   

\author{Zahera Jabeen} \affiliation{Institute
  of Mathematical Sciences, Chennai, India.}  
  \email{zahera@imsc.res.in}
  \author{Neelima Gupte}
  
\affiliation{Department of Physics, Indian Institute of Technology - Madras, Chennai, India}
\email{gupte@physics.iitm.ac.in}
\date{\today} 
\begin{abstract}

Studies of the phase diagram of the coupled sine circle map lattice have identified the presence of two distinct
universality classes of spatiotemporal intermittency 
viz. 
 spatiotemporal
intermittency of the directed percolation  class with a complete set of directed percolation exponents, and spatial
intermittency which does not belong to this class.
We show that these two types of behavior are special cases of a
spreading regime where each site can infect its neighbors permitting an initial disturbance to spread, and a non-spreading regime where no infection is possible , with the two regimes being separated by a line, the infection line. 
The coupled map lattice can be mapped on to an equivalent cellular automaton which shows a  transition from a probabilistic cellular automaton  to a deterministic cellular automaton at the infection
line. 
The origins of the  spreading-non-spreading transition in the coupled map lattice,  as well as the
probabilistic to deterministic transition in the cellular automaton
lie in a dynamical phenomenon, 
an attractor-widening crisis  at the infection line.  
Indications of unstable dimension variability  are  seen in the neighborhood
of the infection line. This may provide useful pointers to the spreading behavior seen in other extended systems. 

\end {abstract}
\maketitle
\section{Introduction}

The identification of the universality class of spatiotemporal
intermittency \cite{sti_papers} in spatially extended systems has been a long standing
problem in the literature. Early conjectures argued that the transition
to spatiotemporal intermittency is a second order phase transition, and
the transition falls in the same universality class as directed
percolation \cite{pomeau}.
This conjecture has
become the central issue in a long-standing debate
\cite{chate,grassberger,bohr,rolf,rupp,takeuchi}, which is still not
completely resolved.

Studies of the coupled sine circle map lattice have
thrown up a number  of intriguing observations of relevance to this
problem \cite{janaki,zjngpre72,zjngpre74}.
This system has regimes of spatiotemporal intermittency (STI) with critical
exponents which fall in the same universality class as directed
percolation(DP), as well as regimes of spatial intermittency (SI) which do not
belong to the DP class. Both these regimes lie on the bifurcation
boundaries of the spatiotemporally fixed point solutions of the map. The
spatiotemporally intermittent regime seen here has an absorbing laminar
state, i.e. a laminar site remains
laminar unless infected by a neighboring turbulent site. The burst
states spread and can percolate through the entire lattice. The system
shows a convincing set of DP exponents in this regime \cite{janaki,zjngpre72,zjngpre74}.
In the spatially intermittent regime, the laminar sites are frozen in
time and the burst sites show temporally periodic or quasi-periodic
behavior. The laminar sites do not get infected by neighboring
turbulent sites.
Hence, the spatially intermittent state is non-spreading and does not
show DP exponents. Thus, both DP and non-DP behavior can be seen for
different parameter regimes of the same system.

In the present paper, we show that the infective DP behavior of
spatiotemporal intermittency and the non-infective behavior of spatial
intermittency are special cases of the more general spreading to
non-spreading transition seen in this system. The spreading and
non-spreading regimes are separated by a line which we call the
infection line. Above the infection line, the burst states can infect
neighboring laminar states and spread through the lattice, whereas
below this line the burst states cannot infect their neighbors and the
non-spreading regime is seen. The infection line intersects the
bifurcation boundary of the synchronized solutions. Intermittent
solutions are seen along this boundary, with the DP type of STI being seen above the
infection  line, and the non-DP SI being seen below the infection
line. Spreading and non-spreading solutions are also seen off the
bifurcation boundary. However, the distribution of laminar lengths shows power-law scaling
only for parameter values which are very close to the bifurcation
boundary, and falls off exponentially as the
parameter values get more distant from the bifurcation boundary.
Other exponents associated with DP behavior are also observed only along the
bifurcation boundaries.

Further insights into the spreading to non-spreading transition are
obtained by mapping the coupled map lattice (CML) onto a cellular automaton (CA). The spreading to
non-spreading transition seen across the infection line maps
on to a transition from a probabilistic cellular automaton to a
deterministic cellular automaton.
The dynamical origins of this transition lie in an attractor widening crisis 
which occurs at the infection line. The existence of this crisis can be inferred from the bifurcation diagram of the system. The finite-time Lyapunov exponents of the system fluctuate in the neighborhood of the infection line indicating the presence of unstable dimension variability. Thus the statistical characterizers of the system show signatures of a dynamical phenomenon. This result could have implications in a wider context.

The paper is organized as follows. In section II, we give details of the coupled sine circle map lattice and the associated phase diagram and discuss the special behavior near the bifurcation boundary in these two regimes. namely spatiotemporal intermittency and spatial intermittency seen in the  spreading  and non-spreading regimes respectively. 
The signatures of the spreading and non-spreading regimes are also seen in the mapping of the coupled map lattice to a cellular automaton and its mean-field analysis. This is discussed in section III. In section IV, we discuss the dynamic origins of the two regimes, namely, the presence of an attractor-widening crisis and unstable dimension variability, as indicated by the fluctuating Lyapunov exponents at the infection line. 
We conclude with a discussion of the implications of our results.

\section{The model and the universality classes}

\begin{figure}
\includegraphics[width=8cm,height=5cm]{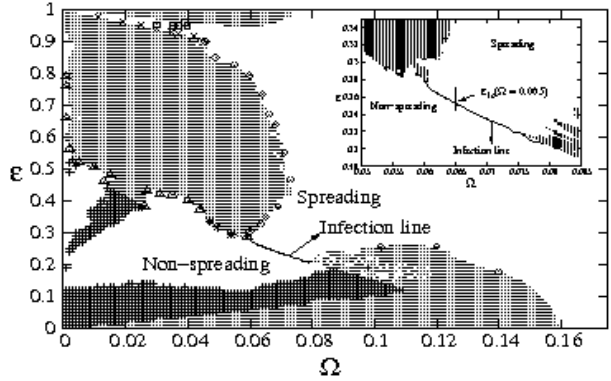}
\caption{ shows the phase diagram of the coupled sine circle map lattice obtained with random initial conditions. The spreading and non-spreading regimes, separated by the infection line have been marked. Spatiotemporal intermittency of the directed percolation class is indicated by diamonds ($\Diamond$), whereas spatial intermittency has been marked with asterisks ($\ast$) and triangles ($\triangle$). \label{pdia}}
\end{figure}

The coupled sine circle map lattice studied here is known to model the
mode-locking behavior \cite{gauri2} seen in coupled
oscillators, Josephson Junction arrays etc.
The model is defined by
the evolution equation
\beq
x_i^{t+1}=(1-\epsilon)f(x_i^t)+\frac{\epsilon}{2}[ f(x_{i-1}^t) +
f(x_{i+1}^t) ]\pmod{1}
\label{evol}
\eeq

where $i$ and $t$ are the discrete site and time indices respectively
and $\epsilon$ is the strength of the coupling between the site $i$ and
its two nearest neighbors. The local on-site map, $f(x)$ is the sine
circle map defined as
$f(x)=x+\Omega-\frac{K}{2\pi}\sin(2\pi x)$,
where, $K$ is the strength of the nonlinearity and $\Omega$ is the
winding number of the  single sine circle map in the absence of the
nonlinearity. We study the system with periodic boundary conditions in
the parameter regime  $0 < \Omega < \frac{1}{2\pi}$ (where
the single circle map has temporal period
1 solutions), $0 < \epsilon< 1$ and $K=1.0$.
The phase diagram of this model evolved with random initial conditions
is shown in Fig. \ref{pdia}.

\begin{figure}
\begin{tabular}{lll}
\includegraphics[width=6cm,height=5cm]{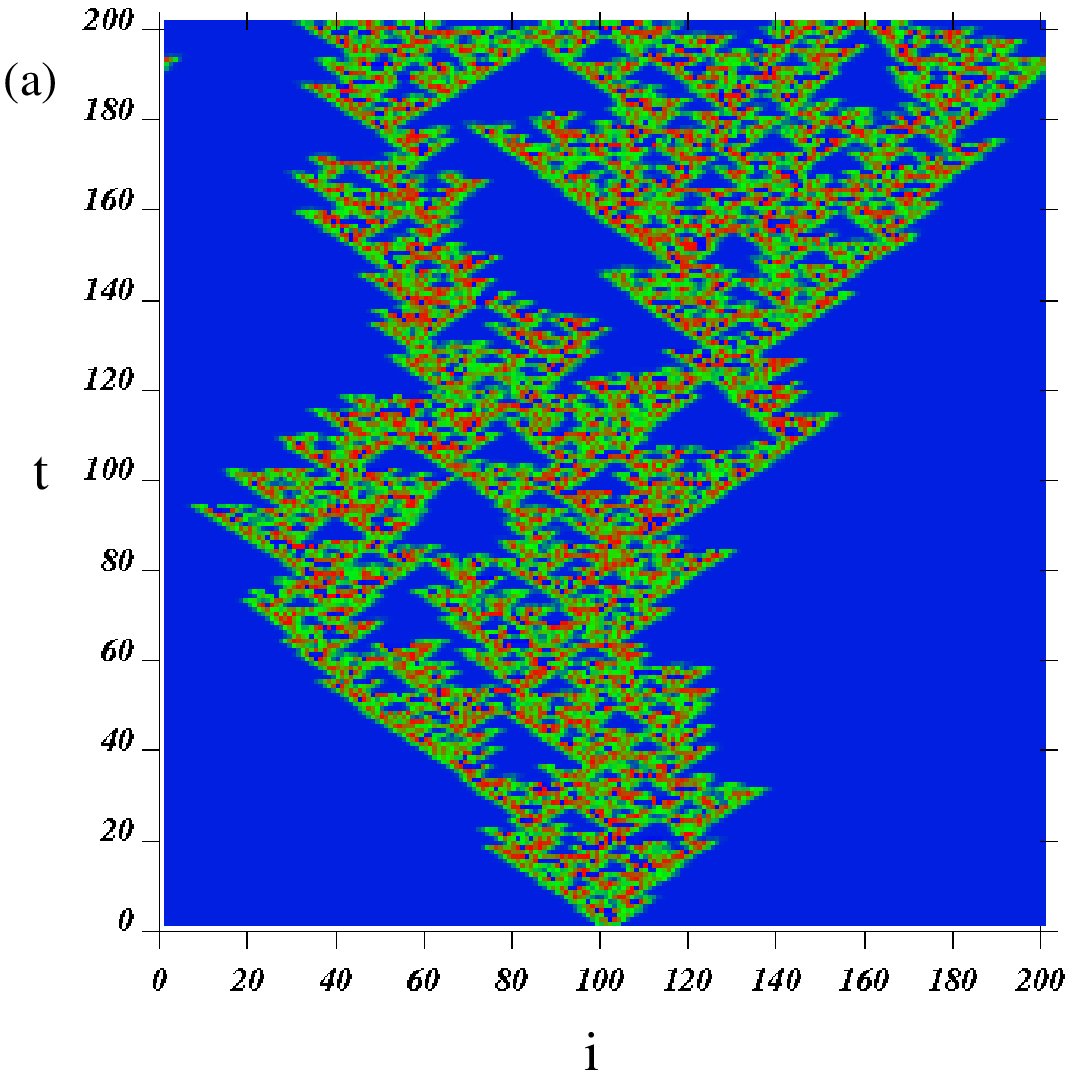} & \includegraphics[width=6cm,height=5cm]{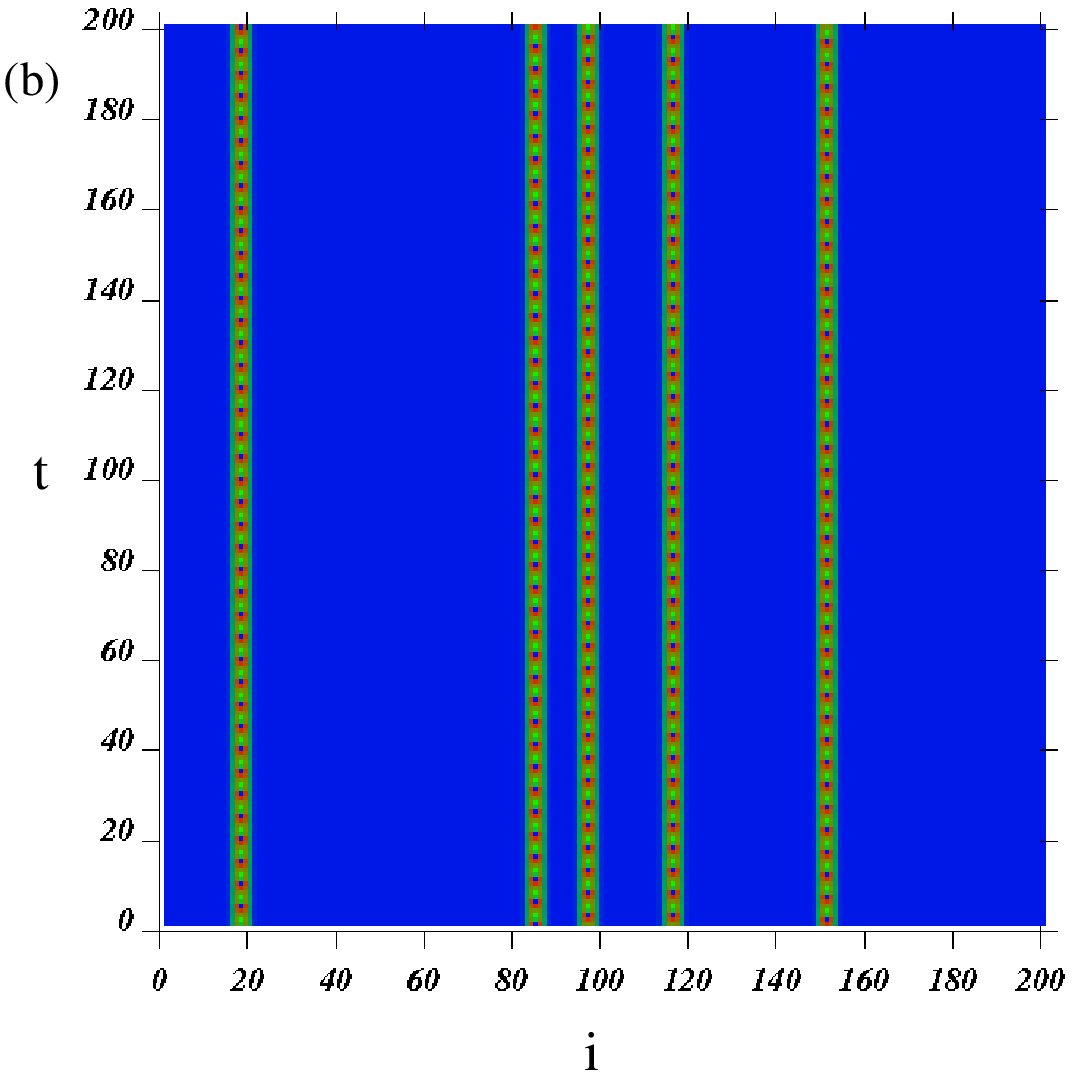}&
\includegraphics[width=1cm,height=5cm]{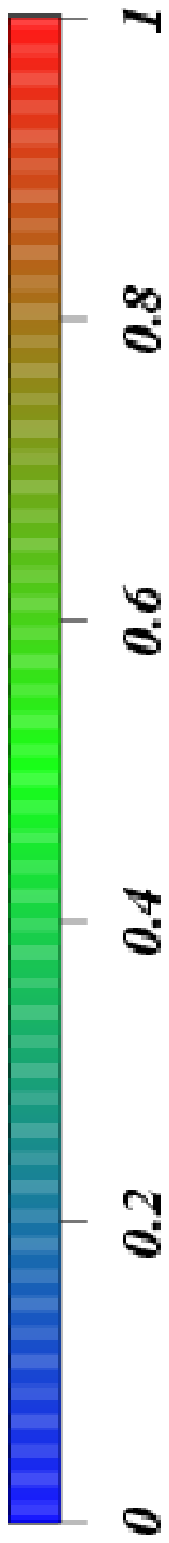}\\
\end{tabular}
\caption{ (Color online) shows the space-time plot of (a) spatiotemporal intermittency seen at $\Omega=0.06, \epsilon=0.7928$ (one of the points marked with diamonds in Fig. \ref{pdia}), and (b) spatial intermittency seen at $\Omega=0.047,\epsilon=0.336$ (marked with asterisk in Fig. \ref{pdia}).  The index $i$ represents the lattice sites and the index $t$ represents time. \label{stplot}}
\end{figure}

Spatiotemporally fixed point solutions, in which all the sites relax to the fixed point $x^{\star}=\frac{1}{2\pi}\sin^{-1}(\frac{2\pi\Omega}{K})$, are seen in a large region of the phase diagram, and are indicated by dots.  Two distinct regimes, separated by a line, which we call the infection line, can be identified in the phase diagram. A spreading regime is seen above the infection line, in which the burst states can infect their neighboring laminar states and spread through the lattice. The spreading nature of the bursts is strongly evident when a disturbance introduced in a laminar background eventually spreads to the entire lattice by infecting the neighboring laminar sites (Fig. \ref{stplot}(a)). This is in contrast to the non-spreading regime seen below the infection line, where the random initial conditions die down to bursts which are localized, and do not infect their neighboring laminar states (Fig. \ref{stplot}(b)). 

In both these regions, spatiotemporal intermittency with co-existing laminar states exhibiting regular temporal behavior, and turbulent states with irregular temporal behavior, can be identified near the bifurcation boundary of spatiotemporally fixed point solutions in the phase diagram.
The spatiotemporal intermittency seen in the spreading regime has been shown to belong to the directed percolation class. In the non-spreading regime, a different class of intermittency called the spatial intermittency is seen near the bifurcation boundary. We discuss these two classes of intermittency briefly in the following sections.

\subsection{Spatiotemporal intermittency of the DP class and the spreading regime}

\begin{table*}[!t]
\begin{center}
\begin{tabular}{c|cc|ccccc|ccccc}
\hline
&\multirow{2}{*}{~ $\Omega$~}&\multirow{2}{*}{ ~$\epsilon_c(\Omega)$~}&\multicolumn{5}{c}{ Static exponents }	&\multicolumn{3}{|c}{ Spreading Exponents}\\
\cline{4-11}
& & &  $z$ &  $\beta/{\nu z} $ &  $\beta$ &   $\eta'$ & { $\zeta$} & $\eta$	&$\delta$	&$z_s$\\
\hline
\multirow{3}{*}{STI-DP}& 0.060	&0.7928	&1.59(0.02)	&0.17(0.02)	&0.293		&1.51(0.01)	&1.68(0.01)&0.315(0.007) &0.16(0.01)	&1.26(0.01)	\\

&0.065	&0.34949	&1.59(0.03)	&0.16(0.01)	&0.273		&1.50(0.01)&1.66(0.01)&0.303(0.001) &0.16(0.01)	&1.27(0.01)	\\
&\multicolumn{2}{c|}{\bf DP }& {\bf 1.58}	& {\bf 0.16}	& {\bf 0.28}	& {\bf 1.51}	&{\bf 1.67} & {\bf 0.313}	&{\bf 0.16}	&{\bf 1.26}		\\
 \hline
\multirow{2}{*}{SI }& 0.04 & 0.402 & - & -  & - & - & 1.10(0.04) & - & - & -\\
& 0.047& 0.336 & -&- & -&- & 1.13(0.02)& -&- &- \\
\hline
\end{tabular}
\caption{ The static and dynamic (spreading) exponents (see \cite{janaki,zjngpre72} for
definitions) obtained at two of the directed
percolation (DP)
points ($\Diamond$-s) in Figure \ref{pdia}
are shown in the first two rows of the table.  The universal DP
exponents are listed in
the third
row.
The laminar length distribution exponent,
$\zeta$, calculated for spatial intermittency (SI) at two  points
marked by  ($\triangle$-s (quasi-periodic bursts) and ($\ast$-s (periodic
bursts) respectively in
Fig. \ref{pdia}, are also listed. The error-bars are shown in the brackets.
The data has been obtained for a lattice of size, $N=10^3$ and has been averaged over $10^3$ initial conditions. The laminar length distributions have been calculated for a lattice of size $N=10^4$ and averaged over $50$ initial conditions.
  \label{dptable}}
\end{center}
\end{table*}

Spatiotemporal intermittency belonging to the directed percolation class is seen in the vicinity of the bifurcation boundary of the spatiotemporally fixed point solutions in the spreading regime \cite{janaki, zjngpre72,zjngpre74}. Some of the  points that show this kind of spatiotemporal intermittency have been marked with diamonds ($\Diamond$) in the phase diagram (Fig. \ref{pdia}). In this type of intermittency, the fixed point solution of the single sine circle map, $x^{\star}$ acts as the laminar state, whereas the burst states lie in the $[0,1]$ interval. Since burst states cannot be spontaneously created, the laminar state acts as an absorbing state. The burst states can either 'percolate' through the lattice by infecting a neighboring laminar state, or die down to  the laminar state. With time acting as the directed axis, the analogy with directed percolation is complete. Moreover, this class of spatiotemporal intermittency is remarkably free of coherent structures, some times called 'solitons', which spoilt the analogy with directed percolation in other models such as the Chat\'e Manneville coupled map lattice \cite{chate}.

An entire set of static and dynamic scaling exponents was obtained in this parameter regime, which showed good agreement with the universal DP exponents. The exponents obtained, after averaging over $10^3$ initial conditions, at two such parameter values are listed in Table \ref{dptable}. Similar exponents have been seen at points marked with diamonds ($\Diamond$) in the phase diagram (Figure \ref{pdia}).  These exponents also satisfy the hyperscaling relations for the static exponents ($2\beta/\nu=d-2+\eta'$), and the spreading exponents ($4\delta+2\eta=d z_s$) for $d=1$. Hence, this class of spatiotemporal intermittency seen at the bifurcation boundary in the spreading regime belongs convincingly to the directed percolation class. These scaling exponents are however, seen only near the bifurcation boundary.  
The distribution of laminar lengths
 for the spreading solutions  off the bifurcation boundary,
show an exponential fall off. 
Thus STI of the DP class is a special case of the spatiotemporal behavior seen in the spreading regime.


A different class of intermittency is seen below the infection line, i.e in the non-spreading regime. This is the phenomenon of spatial intermittency.

\subsection{Spatial intermittency and the non-spreading regime}

Below the infection line, in the non-spreading
regime,
the laminar sites are
the synchronized fixed point $x^*$ and the burst sites are either temporally
frozen, periodic or aperiodic.
These bursts are  very stable and do not die down with time.

In the vicinity of the bifurcation boundary, a special class of intermittency called spatial intermittency is seen. In this case, the temporal behavior of the burst states is either quasi-periodic or periodic. Some of these points have been marked with triangles ($\triangle$, quasi-periodic)  and asterisks ($\ast$, periodic) in the phase diagram (Fig. \ref{pdia}). In this class of intermittency, the distribution of laminar lengths, obtained by averaging over different random initial conditions, scales as a power-law of the form $P(l)\sim l^{-\zeta}$, with $\zeta$ as the associated scaling exponent \cite{zjngpre74}.  The exponent $\zeta$ obtained in this case was found to be $\sim1.1$ (Table \ref{dptable}). A similar scaling exponent has been seen in the case of the spatial intermittency observed in  inhomogeneous logistic map lattice \cite{ashutosh}. 
The laminar length distributions  seen for other non-spreading
solutions at points off
the bifurcation boundary show an exponential
decrease. Thus the spatially intermittent solutions are special cases 
of the solutions in the non-spreading regime. 

Thus the sine circle coupled map lattice shows a  transition from a spreading regime
to a non-spreading regime at the infection line. In order to gain further insights into this
transition,
we  map the coupled map lattice to a stochastic model,
a probabilistic cellular automaton of the Domany-Kinzel type \cite{chate,domanykinzel}.

\section{Mapping to an equivalent cellular automaton}

\begin{table}[!b]
\begin{center}
\begin{tabular}{ccccccccc}
\hline
&$\Omega$&$\epsilon$ & $p_0$ & $p_1$ & $p_2$ & $p_3$ & $p_4$ & $p_5$\\
\hline
\multirow{2}{*}{S(DP)}&~0.060~ & ~0.7928~ &~0.0~ &~0.220~ &~0.0~ &~0.933~ &~0.627~ &~0.984~\\
&~0.073~ &~0.4664~ &~0.0~ &~0.150~ &~0.0~ &~0.938~ &~0.439~ &~0.993~\\
\hline

~\multirow{2}{*}{S}~&~\multirow{2}{*}{0.070}~&~0.264~ &  ~0.0~ &  ~0.140~ &  ~0.0~ &  ~0.982~ &  ~0.391~ &  ~0.999~ \\

&&~0.248~ &  ~0.0~ &  ~0.050~ &  ~0.0~ &  ~0.989~ &  ~0.160~ &  ~0.999~ \\
\hline
~\multirow{2}{*}{NS}~&~\multirow{2}{*}{0.070}~& ~0.232~ &  ~0.0~ &  ~0.000~ & ~0.0~ &  ~1.000~ &  ~0.000~ & ~1.000~ \\
&& ~0.228~ &  ~0.0~ &  ~0.000~ & ~0.0~ &  ~1.000~ &  ~0.000~ & ~1.000~ \\
\hline

\multirow{2}{*}{NS(SI)}&~0.031~& ~0.420~ & ~0.0~ & ~0.000~& ~0.0~ &1.000~&~0.000~&~1.000~\\
& ~0.044~ & ~0.373~ & ~0.0~ & ~0.000~& ~0.0~ &1.000~&~0.000~&~1.000~\\
\hline

\end{tabular}
\end{center}
\caption{shows the probabilities $p_k$'s obtained in the spreading (S), non-spreading (NS) regimes and at directed percolation (DP) and spatial intermittency (SI) points.\label{pkdpsi}}
\end{table}

Signatures of the spreading to non-spreading transition at the infection line are seen in a mapping of the coupled map lattice to a stochastic model namely, a probabilistic cellular automaton of the Domany-Kinzel type \cite{chate, domanykinzel}. The equivalent cellular automaton, set up to mimic the dynamics of the laminar and burst
states in the coupled map lattice  defined in eq. \ref{evol}, is defined on a one dimensional lattice
of size $N$. The state variable $v_i^t$ at site $i$ and at time $t$ is assigned the value $v_i^t=0$ if the site is in the laminar state, and $v_i^t=1$ if the site is in the burst state.

As in the coupled map lattice equation \ref{evol}, the
probability of the site $i$ at time $t+1$ being in the burst state 
depends on the state of the sites $i-1, i$ and $i+1$ at time $t$. 
We therefore define the cellular automaton dynamics in this system
by the conditional probability $P(v_i^{t+1}|v_{i-1}^t,v_i^t,v_{i+1}^t)$.
There are $2^3$ possible configurations and 
the symmetry between
the sites $i-1$ and $i+1$ in equation \ref{evol} gives the effective probabilities $p_k$'s as  $p_0=P(1|000)$ , $p_1=P(1|001)=P(1|100)$, $p_2=P(1|010)$, $p_3=P(1|011)=P(1|110)$, $p_4=P(1|101)$ and $p_5=P(1|111)$. These probabilities then define the update rules of the cellular automaton.

We estimate these probabilities $p_k$, for a given set of parameter values, from the numerical evolution 
of the coupled map lattice with random initial conditions.
The probabilities $p_k$'s are calculated by finding  
the fraction of sites $i$ which exist in the burst state $v_i^{t+1}=1$ at time $t+1$, given
that the site $i$ and its nearest neighbors $i-1$ and
$i+1$ existed in state $k$ at time $t$. 
That is, the probability $p_k$ is determined using $p_k=\frac{N_1^{k}}{N_0^k+N_1^k}$, where
 $N_0^k$ and $N_1^k$ are the number of
sites,  which at time $t$ were the central sites of the configuration $k$, and at time  $t+1$ exist in the laminar states ($v_i^{t+1}=0$) and the burst states ($v_i^{t+1}=1$) respectively. 
These probabilities, were extracted from a  coupled map lattice of size $N=10^3$
averaged over $14000$ time steps discarding  a transient of $1000$
time steps, and averaged over $200$ initial conditions. The probabilities $p_k$ obtained in the spreading and
non-spreading regimes in the phase diagram are listed in 
Table \ref{pkdpsi}. 

\begin{figure*}[!t]
\begin{center}
\begin{tabular}{cc}
\includegraphics[height=1.9in,width=2.8in]{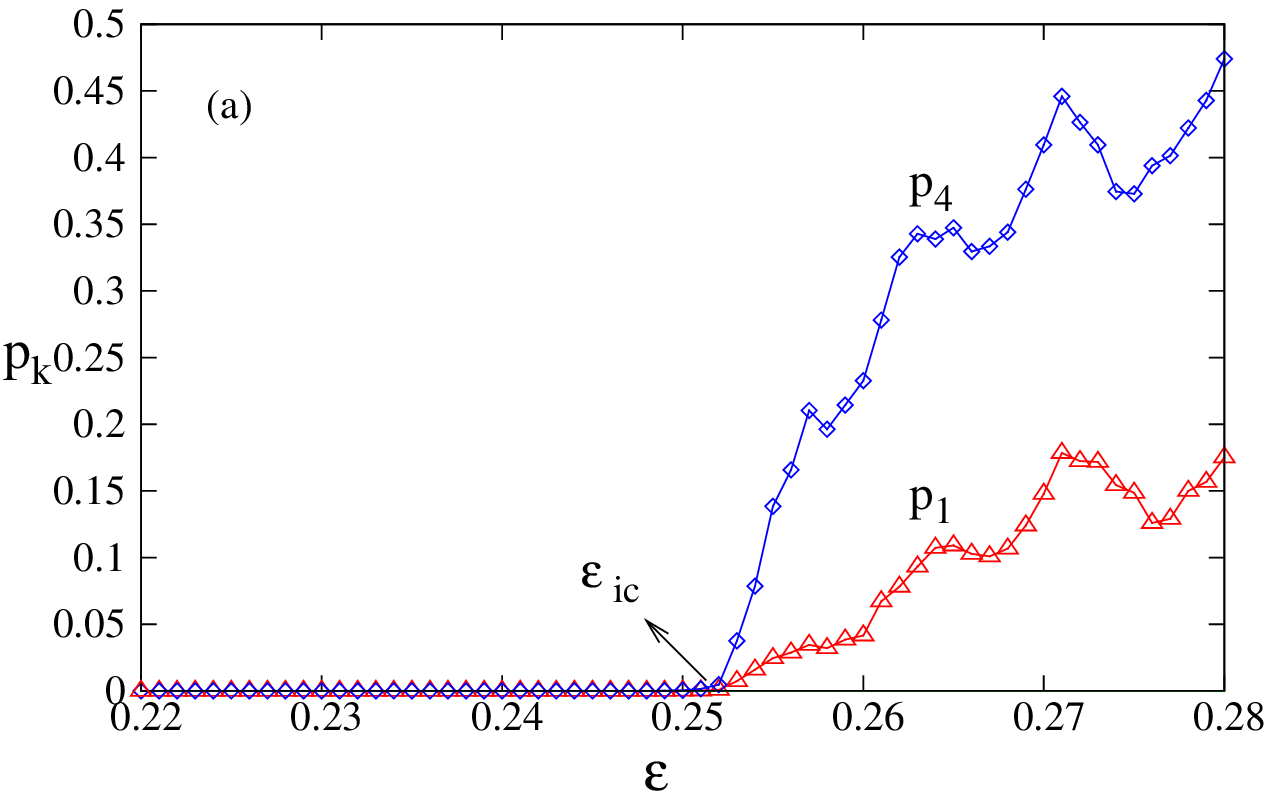} & \includegraphics[height=1.9in,width=2.8in]{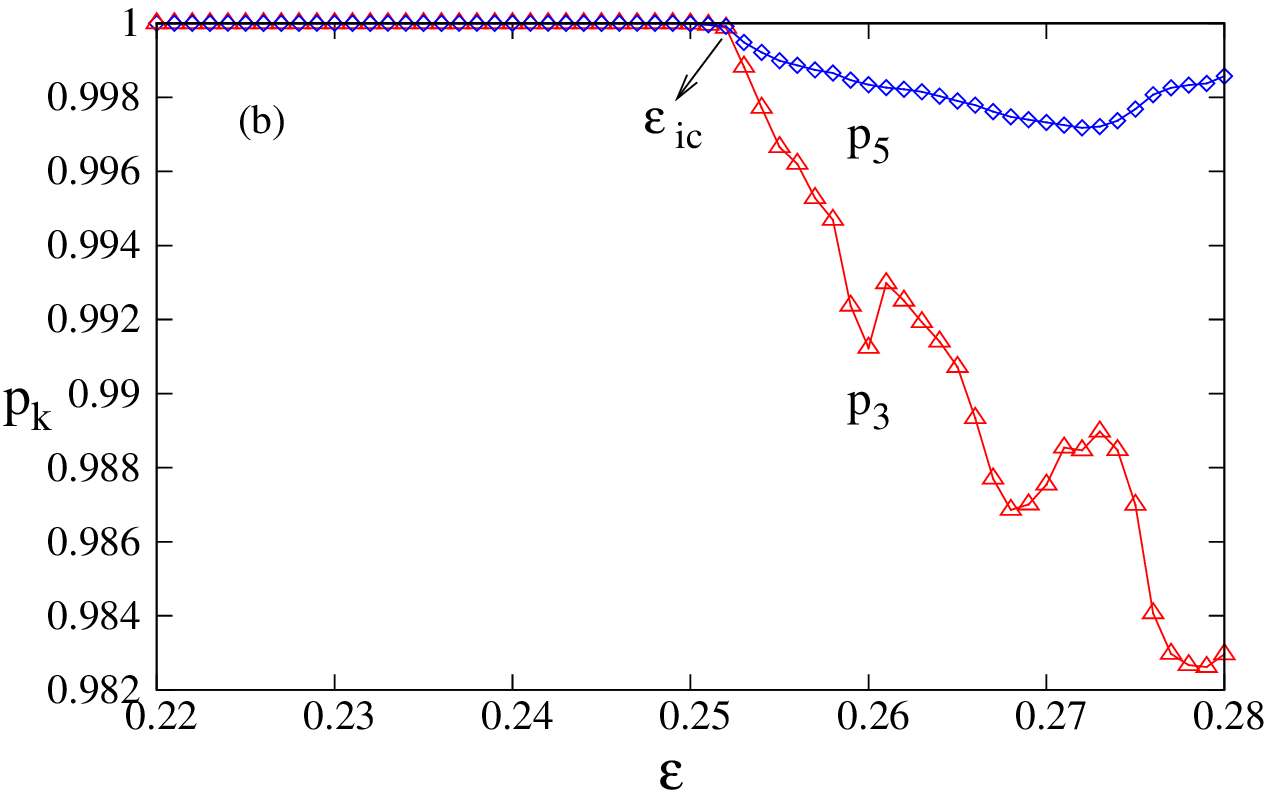}\\ 
\end{tabular}
\end{center}
\caption{ (Color online) shows the probabilities associated with the update rules $p_1, p_3, p_4, p_5$ plotted as a function of the coupling strength, $\epsilon$ at $\Omega=0.065$. In the non-spreading regime ($\epsilon<\epsilon_{ic}$), the probabilities are either $0$ or $1$, whereas they lie in the interval $(0,1)$ in the spreading regime ($\epsilon\geq\epsilon_{ic}$). \label{probfig}}
\end{figure*}

It is clear that the condition for an absorbing state is realized in the probability $p_0=P(1|000)$ which is seen to be zero in both regimes. The probabilities
$p_1=P(1|001)=P(1|100)$ and  $p_4=P(1|101)$ essentially define infection
probabilities, which estimate the probability of a laminar site being infected by its burst neighbor
or neighbors, to change to a burst site.

 We can see from Table \ref{pkdpsi} that these probabilities show
drastically different behavior in the spreading and non-spreading
regimes. 
In the case of the spreading regime, the probabilities obtained are seen to lie 
in the open interval $(0,1)$. Therefore, the dynamics in the spreading regime is described by 
 a probabilistic cellular automaton (PCA) wherein the cellular automaton rules are
probabilistic in nature. In contrast, in the case of the non-spreading regime, 
 in addition to $p_0$ and $p_2$ which are zero for the STI of the DP type \cite{notep2}, the infection
probabilities  $p_1$ and $p_4$ are also seen to be equal to zero. Hence, the infection probabilities characterize the non-infective nature of the bursts. Moreover, the probabilities $p_3=P(1|011)=P(1|110)$ and $p_5=P(1|111)$ take the value $1$, which indicates the robustness of the burst states. We note that the probabilities obtained here only take the values $0$ or $1$. Hence, we obtain a deterministic cellular automaton (DCA) in the non-spreading regime, wherein given a state $k$ at time $t$, the state of the site $i$ at time $t+1$ is decisively known with probability zero or one. This is further illustrated in Figure \ref{probfig}, where the probabilities have been plotted as a function of the coupling strength $\epsilon$ at $\Omega=0.065$. The probabilities show a distinct change to values other than zero or one, at the point where spreading regime starts viz. $\epsilon>\epsilon_{ic}$ (See inset in Fig. 1). Thus a transition from a probabilistic to a deterministic cellular automaton is seen at the infection line.

It is therefore clear that the  spatiotemporal intermittency of the directed percolation class seen in the spreading regime can be modeled by a PCA, whereas spatial intermittency observed in the non-spreading regime can be modeled by a DCA. In the following section, we show that the probabilistic cellular automata obtained at the STI of the DP points indeed mimics the dynamics of the coupled map lattice and shows scaling exponents matching with the universal directed percolation exponents.

\subsection{Probabilistic cellular automaton and directed percolation}

As mentioned previously, the dynamics seen in the case of spatiotemporal intermittency in the spreading regime resembles that of directed percolation. The stochastic nature of the burst states is further emphasized in the probabilistic 
cellular automaton obtained in this regime. In this section, we show that the probabilistic cellular automata obtained at points in the phase diagram which show DP-like spatiotemporal intermittency, indeed show scaling behavior which match with the directed percolation class.

\begin{figure*}[!t]
\begin{center}
\begin{tabular}{cc}
\includegraphics[scale=0.55]{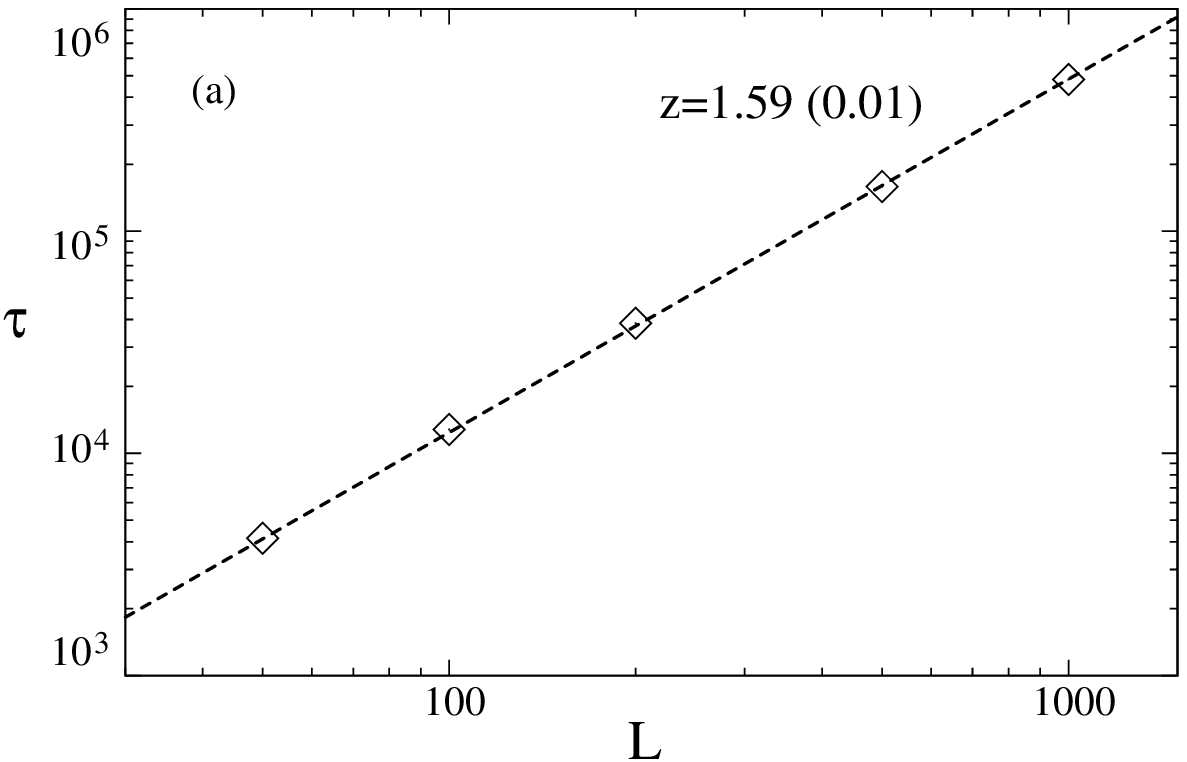} & \includegraphics[scale=0.55]{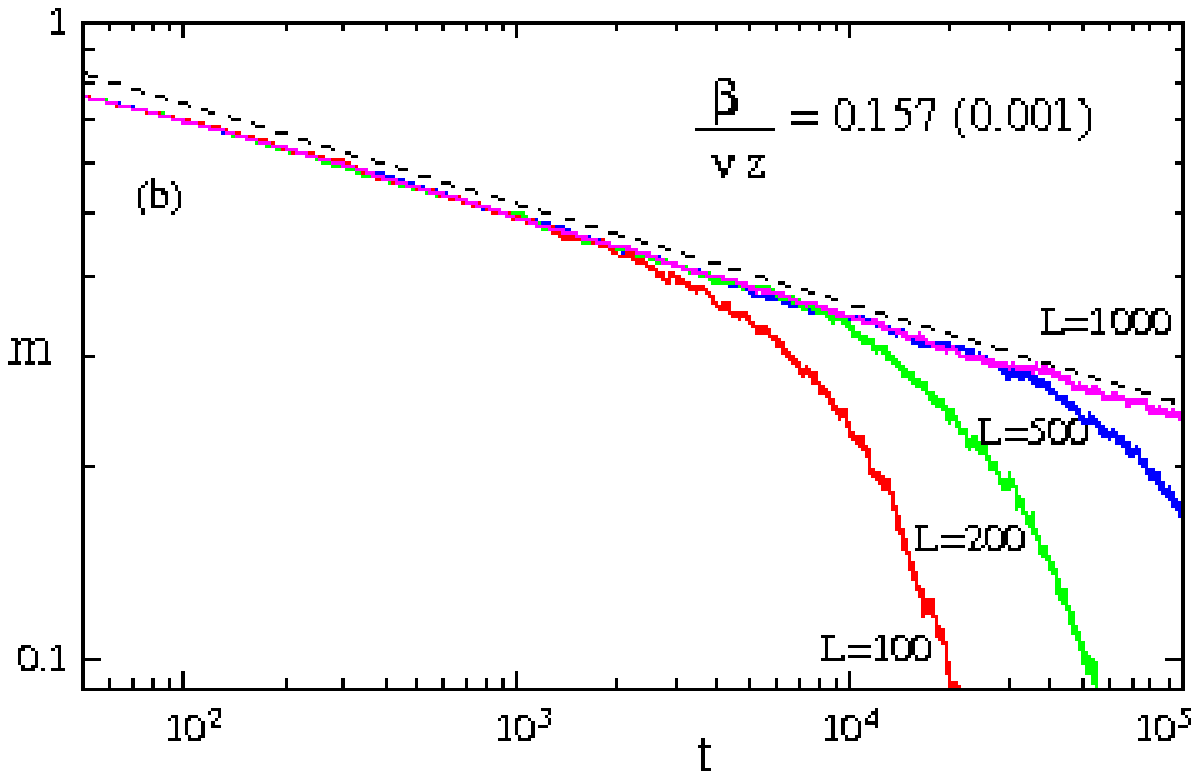}\\ 
{\includegraphics[scale=0.55]{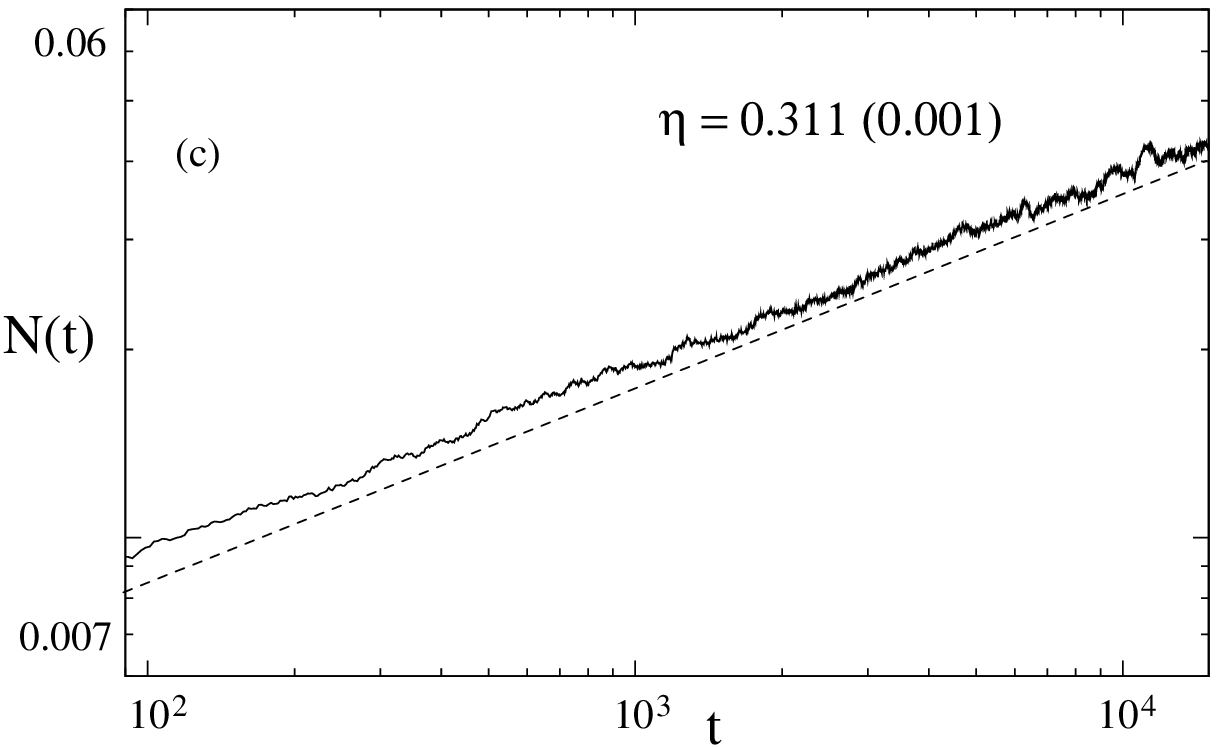}}& \includegraphics[scale=0.55]{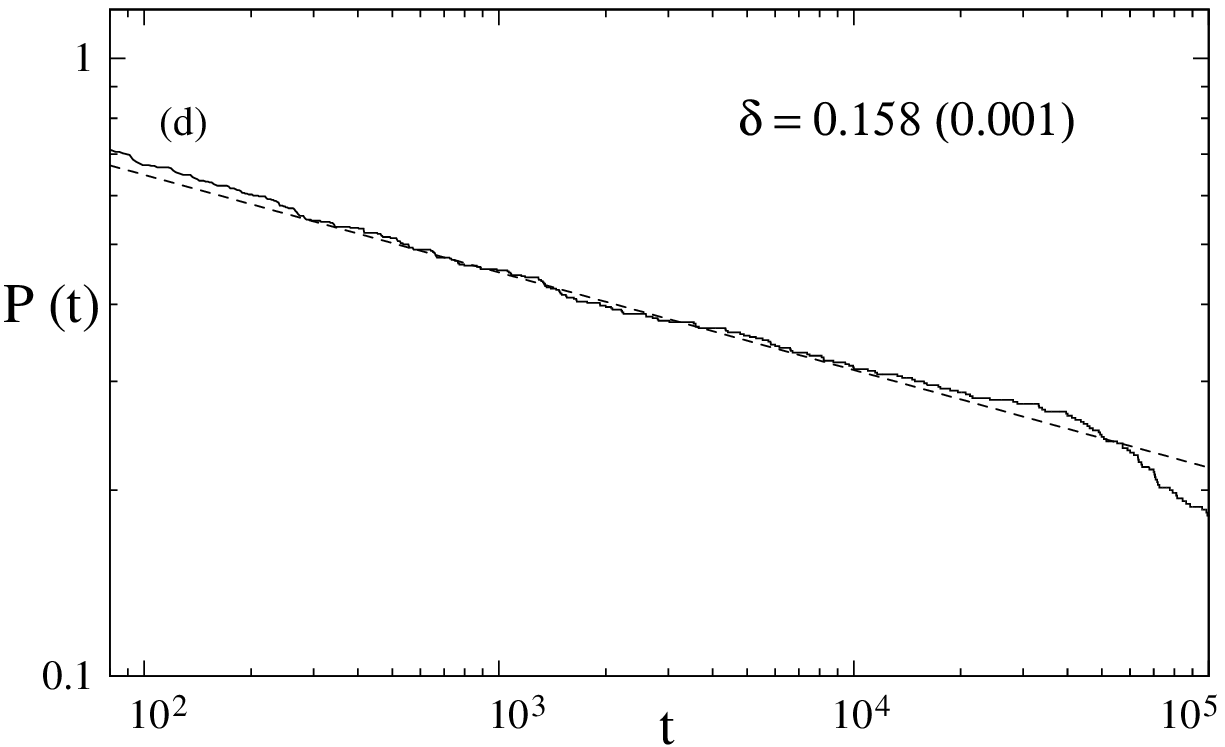}\\
\end{tabular}
\end{center}
\caption{(Color online) shows the log-log (base 10) plot of physical quantities calculated for the probabilistic cellular automaton associated with the DP point $\Omega=0.06, \epsilon=0.7928$. The figure shows the (a) escape time $\tau$ vs length of the lattice $L$, (b) order parameter $m$ vs time $t$, and the spreading properties (c) the fraction of burst sites $N(t)$ vs $t$, and (d) the survival probability $P(t)$ plotted as a function of time $t$. The fits to the plots and the corresponding scaling exponents(error bars in brackets) obtained are also shown. \label{cadp}}
\end{figure*}

Figure \ref{cadp} shows some of the physical quantities associated with the static and dynamical properties of the system. The scaling exponents obtained for these quantities are also shown in the figure. The probabilities associated with the PCA obtained at the parameter value $\Omega=0.06, \epsilon=0.7928$, were found to be $p_0=0.0, p_1=0.213116, p_2=0.000081, p_3=0.902515, p_4=0.569464$ and $p_5=0.967595$.  When the cellular automaton is prepared with random initial conditions, such that the sites take either $0$ or $1$ values with equal probabilities at time $t=0$, the burst state ($1$ state) is seen to die down with time. The relaxation time $\tau$, which determines the time taken by the lattice to relax to a laminar state ($0$ state), scales as a function of the size of the lattice $L$ as $\tau\sim L^z$ (Fig. \ref{cadp}(a)). The order parameter $m$, defined as the fraction of burst sites in the lattice, decreases with time as $m\sim t^{-\beta/{\nu z}}$ (Fig. \ref{cadp}(b)). In contrast, when a pair of burst states are introduced in a lattice prepared in the laminar state at time $t=0$, the burst states are seen to grow with time. The dynamical quantities associated with the growth of the burst states viz. the fraction of burst sites $N(t)$, and the survival probability $P(t)$, which is defined as the fraction of initial conditions which yield a non-zero number of burst states at time $t$, are shown in Fig. \ref{cadp}(c) and (d). The respective  scaling exponents obtained show good agreement with the directed percolation class. Thus the cellular automaton model obtained in the region of spatiotemporal intermittency seen in the spreading regime exhibits scaling behavior similar to the directed percolation class, as expected. 

\begin{figure}[!t]
\begin{center}
\includegraphics[height=5.5cm,width=8cm]{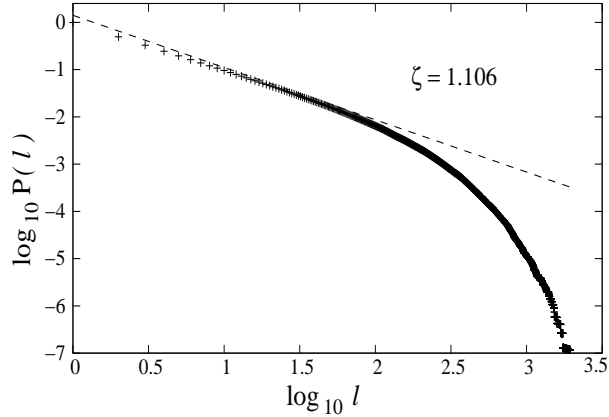}
\end{center}
\caption{shows the laminar length distribution obtained for the deterministic cellular automaton obtained at the spatial intermittency point $\Omega=0.04, \epsilon=0.402$. The exponent obtained is $\zeta=1.106\pm 0.005$.\label{siexp}}
\end{figure}

We note that in the case of the non-spreading regime, the deterministic cellular automaton obtained acts like an identity mapping. Therefore, the scaling behavior in the laminar length distribution depends strongly on the initial condition namely, the fraction of burst states in the lattice. To mimic the dynamics exhibited by the coupled map lattice, we calculate the fraction of burst sites in a coupled map lattice of size $20000$ at $\Omega=0.04, \epsilon=0.402$,  after discarding $60000$ transients and averaging over $50$ random initial conditions. The fraction of burst sites came out to be $p=0.0684$. Fig. \ref{siexp} shows the scaling behavior of the distribution of laminar lengths for the deterministic cellular automaton, obtained for this initial condition, in which bursts were introduced with
probability $p$. The exponent seen in the cellular automaton turns out to have the value $\zeta \sim1.1$ as in the coupled map lattice.

A mean-field study of the cellular automaton gives further insights into the transition from probabilistic to deterministic cellular automaton, as well as into the initial condition dependence of the cellular automata. We discuss this in the subsequent section.

\subsection{Mean-field analysis of the cellular automaton}

We study a mean field approximation of the probabilistic cellular automaton which gives a better understanding of the probabilistic to deterministic cellular automaton transition as well as gives probability bounds for the directed percolation-like behavior \cite{bagnolica}. Let $m_t$ and $m_{t+1}$ be the density of burst states in the lattice at the $t^{th}$ and $t+1^{th}$ time step.  The mean-field for the probabilistic cellular automaton can be defined, according to the cellular automaton rules, as 
\beq
m_{t+1}= (2 p_1 + p_2) m_t(1-m_t)^2 + (2 p_3 +p_4) m_t^2 (1-m_t) + p_5 m_t^3
\eeq
By approximating $p_5=1$ and $p_2=0$, the mean field equation reduces to 
\beq 
m_{t+1}=  2 p_1~  m_t(1-m_t)^2 +  (2 p_3 +p_4)~ m_t^2 (1-m_t) + m_t^3
 \label{mfca}
\eeq. 

\begin{figure}[!t]
\begin{center}
\includegraphics[height=5.5cm,width=8cm]{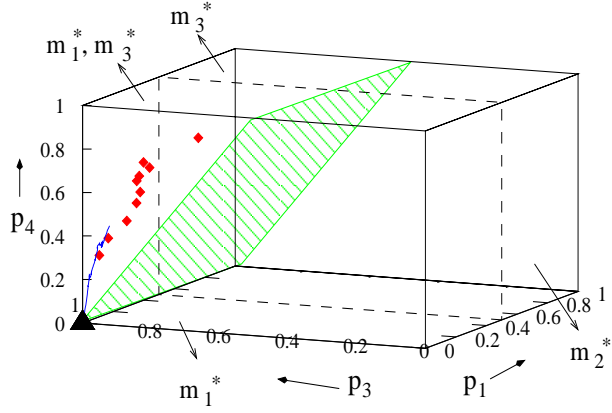}
\end{center}
\caption{ (Color online) shows the stability regions of the three fixed points of the  mean-field equation \ref{mfca}, $m_1^{\star}, m_2^{\star},$ and $m_3^{\star}$, in the $\{p_1,p_3,p_4\}$ phase space. The probabilities of the DP points (marked with diamonds), as well as for points in the spreading regime at $\Omega=0.065$ (dotted line) have been shown. The triangle corresponds to the probabilities in the non-spreading regime. \label{cubemf}}
\end{figure}

The three fixed points of this equation for a set of parameters $\{p_1,p_3,p_4\}$ are $m_1^{\star}=0, m_2^{\star}=\frac{2 p_1-1}{1+2p_1-2p_3-p_4},$ and  $m_3^{\star}=1$. 
A linear stability analysis of the mean-field equation indicates that the fixed point $m_1^{\star}=0$, which corresponds to the absorbing state, is stable  when $p_1<\frac{1}{2}$. The fixed point $m_2^{\star}$ is stable when $2p_3+p_4\leq 2$ and $2p_1 \geq 1$. The fixed point $m_3^{\star}=1$, which corresponds to the completely turbulent state, is stable  when $2p_3+p_4\geq 2$. The stability regions of these three fixed points in the $\{p_1, p_3, p_4\}$ space have been indicated in Fig. \ref{cubemf}. The $2p_3+p_4=2$ plane, above which $m_3^{\star}=1$ solutions are found to be stable, is shown in the figure.

\begin{figure*}[!t]
\begin{center}
\begin{tabular}{cc}
{\includegraphics[scale=0.7]{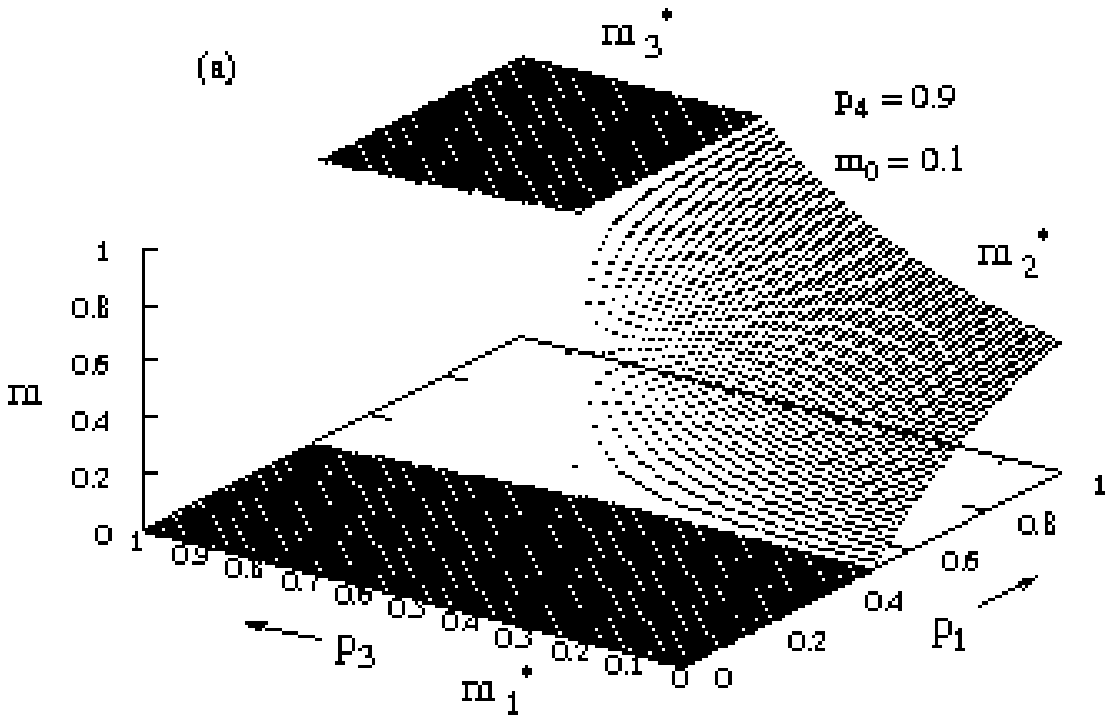}} & {\includegraphics[scale=0.7]{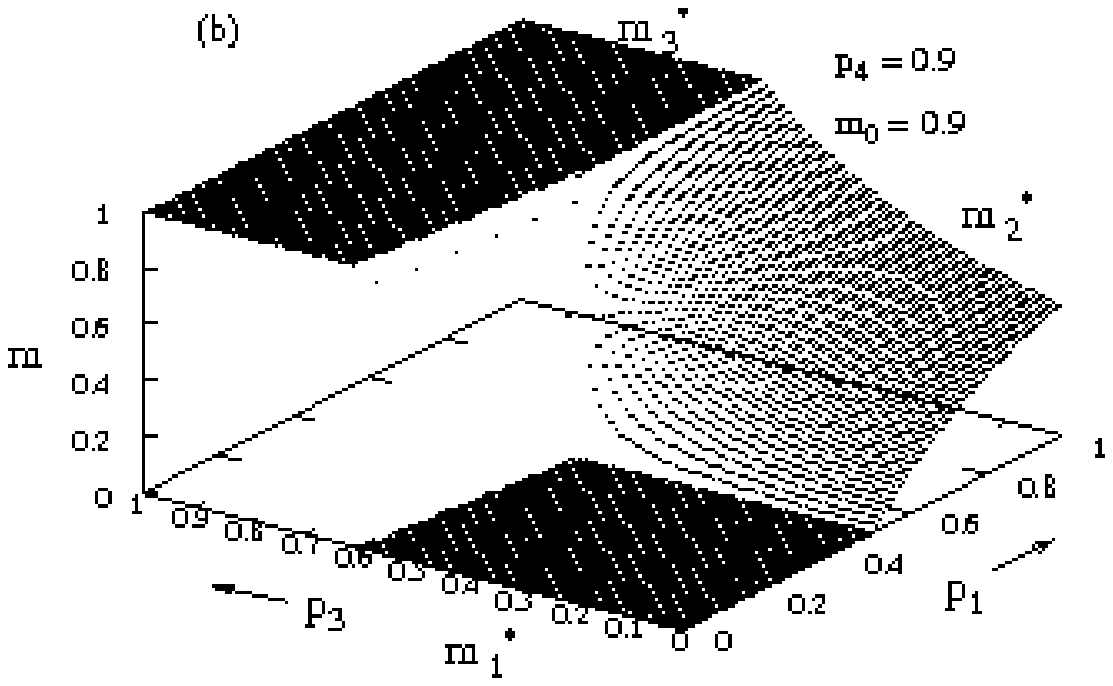}}\\ 
\end{tabular}
\end{center}
\caption{shows the $\{m,p_1,p_3\}$ phase diagram of the mean field map at $p_4=0.9$ with initial $m$ at time $t=0$ chosen as (a) $m_0=0.1$ and (b) $m_0=0.9$. In the co-existence region ($2 p_3\geq 1.1$ and $p_1\leq 0.5$), the density of burst sites $m$, settles down to either $m_1^{\star}$ or $m_3^{\star}$ depending on the initial density $m_0$ chosen.\label{itmap}}
\end{figure*}

We see that both $m_1^{\star}=0$ and $m_3^{\star}=1$ solutions are stable, when $p_1\leq \frac{1}{2}$ and $2p_3+p_4\geq 2$. This is the co-existence region. The system settles down to either the completely laminar state ($m_1^{\star}=0$) or the completely turbulent state ($m_3^{\star}=1$) in the co-existence region, depending on the initial condition $m_t$ at $t=0$.  The solutions settle down to the $m_1^{\star}=0$ solution if initial density, $m_0 \in [0,m_2^{\star})$ whereas, they tend towards $m_3^{\star}=1$ if $m_0\in (m_2^{\star},1]$. For instance, we iterate the map equation \ref{mfca} by choosing $p_4=0.9$ and study the $\{m, p_1, p_3\}$ phase diagram for two initial conditions $m_0=0.1$ and $m_0=0.9$. The $\{m, p_1, p_3\}$ phase diagram obtained is shown in Figure \ref{itmap}(a) and (b) respectively. We see that the solutions settled down to $m^{\star}=0$ in the co-existence region in the first case, whereas they settle down to $m^{\star}=1$ in the latter case. Hence, the solutions in the co-existence region are initial condition dependent.

Interestingly, we find that the probabilities associated with the spreading regime of the coupled map lattice, including those obtained for the directed percolation points in the spreading regime, lie in the co-existence region (Figure \ref{cubemf}). Due to this, the probabilistic cellular automaton has a very strong
initial condition dependence in the spreading region. The correct choice of initial conditions leads 
to the laminar absorbing state. Other choices can end up easily in the $m=1$ state. However, the points at which directed percolation is seen are exceptions to this. The probabilistic cellular automata obtained at these points show
scaling behavior consistent with the directed percolation phenomena irrespective of the choice of initial conditions, as was discussed in
the previous section.

The probabilities associated with the deterministic cellular automaton seen in the non-spreading regime (i.e. $ p_1=0,~p_3=1,~p_4=0 $) lie at the vertex  belonging to the co-existence region in this cube. This point has been marked  with a triangle in Figure \ref{cubemf}. 
As we approach the spatial intermittency points along the bifurcation boundary, the cellular automaton probabilities obtained at the directed percolation points tend towards this vertex. Similarly, the probabilities obtained in the spreading regime tend towards this vertex, as we cross the infection line and enter the non-spreading regime. Therefore, the deterministic cellular automaton is a limiting case of the probabilistic cellular automaton seen in the spreading regime.

This confirms that the presence of both spreading and non-spreading regimes separated by the infection line in the phase diagram, which accounts for the presence of both directed percolation-like spatiotemporal intermittency and spatial intermittency in the phase diagram, is manifested in the form of a transition from probabilistic to deterministic cellular automaton at the infection line. However, the origins of the spreading nature of the burst states above the infection line are not clear. In the next section, we identify the dynamical  origins of the spreading states.

\section{Crisis and unstable dimension variability at the infection line}

It is clear from the preceding discussion that the behavior of the burst states changes drastically at the infection line leading to the spreading to non-spreading transition and the existence of distinct classes of spatiotemporal intermittency in the phase diagram. We now have indications that
 the origin of the spreading states lies in an attractor-widening crisis at the infection line as well as the presence of unstable dimension variability in the vicinity of the infection line, as we will see in this section.

\begin{figure}[!t]
\begin{center}
\includegraphics[scale=0.7]{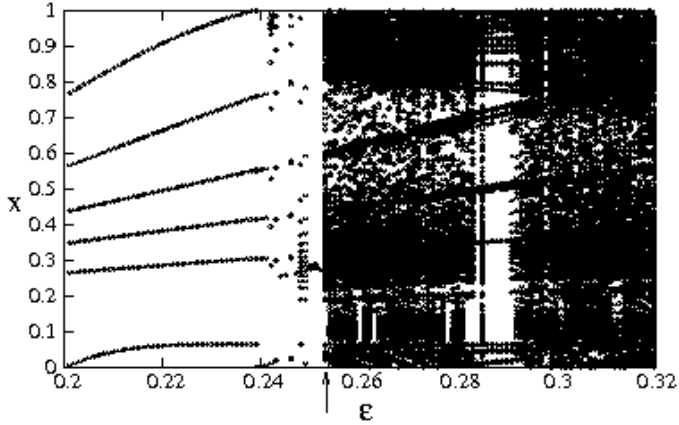}\\
\end{center}
\caption{ shows the bifurcation diagram of the coupled map lattice in which the variable ${x}$ associated with a typical site has been plotted over $500$ time steps, as a function of the coupling strength $\epsilon$ in the neighborhood of the infection line, at $\Omega=0.065$.  \label{crisis}}
\end{figure}

A crisis is said to have occurred in a dynamical system  when a sudden change in the attractor takes place when a system parameter is changed \cite{yorke1}. We observe such a  phenomenon in our coupled map lattice, when we change the strength of the coupling between sites, $\epsilon$, for a given $\Omega$ in the vicinity of the infection line. This has been illustrated in Fig. \ref{crisis}. Here, the variable $x$ associated with a typical site has been plotted over $500$ time steps, as a function of the coupling strength $\epsilon$ at $\Omega=0.065$.

The range of $\epsilon$
values on the vertical axis of Fig. \ref{pdia} cuts across the infection
line at $\epsilon_{ic}=0.252$ for $\Omega=0.065$.  The bifurcation diagram clearly shows that an attractor widening
crisis \cite{greb} appears at this point. Similar behavior is seen for other sites. The spreading regime seen in
the phase diagram emerges exactly at the point at which the attractor
widens, with the non-spreading regime corresponding to the pre-widening
regime. This widening also identifies the point at which the equivalent
cellular automaton undergoes a probabilistic to deterministic transition. In the pre-crisis
region, each site follows either a periodic or quasi-periodic trajectory
and is not infected by the behavior of its neighbors. Thus, its CA analogue
is deterministic as listed in Table \ref{pkdpsi}. In the post-crisis regime,
each site is able to access the full $x$ range, as well as infect its
neighbors, and the bursting and spreading behavior characteristic of the spreading regime is
seen. This is reflected in the equivalent cellular automaton by a transition to probabilistic behavior (Table \ref{pkdpsi}).
It is to be noted that the volume of the attractor in phase space
will be much larger post-crisis, as compared to the pre-crisis volume.

We also see indications of unstable dimension variability in the vicinity of the infection line. An attractor is said to have unstable dimension variability, if it possesses periodic orbits with different number of stable and unstable directions \cite{kost,lai}. A trajectory visiting the neighborhood of these periodic orbits, experiences a fluctuating number of unstable directions, as time progresses. A signature of this phenomenon can be seen in the finite time Lyapunov exponents (FTLE) of the system which fluctuate about zero, as the dynamics evolves. 

\begin{figure}[!t]
\begin{center}
\begin{tabular}{cc}
\includegraphics[scale=0.6]{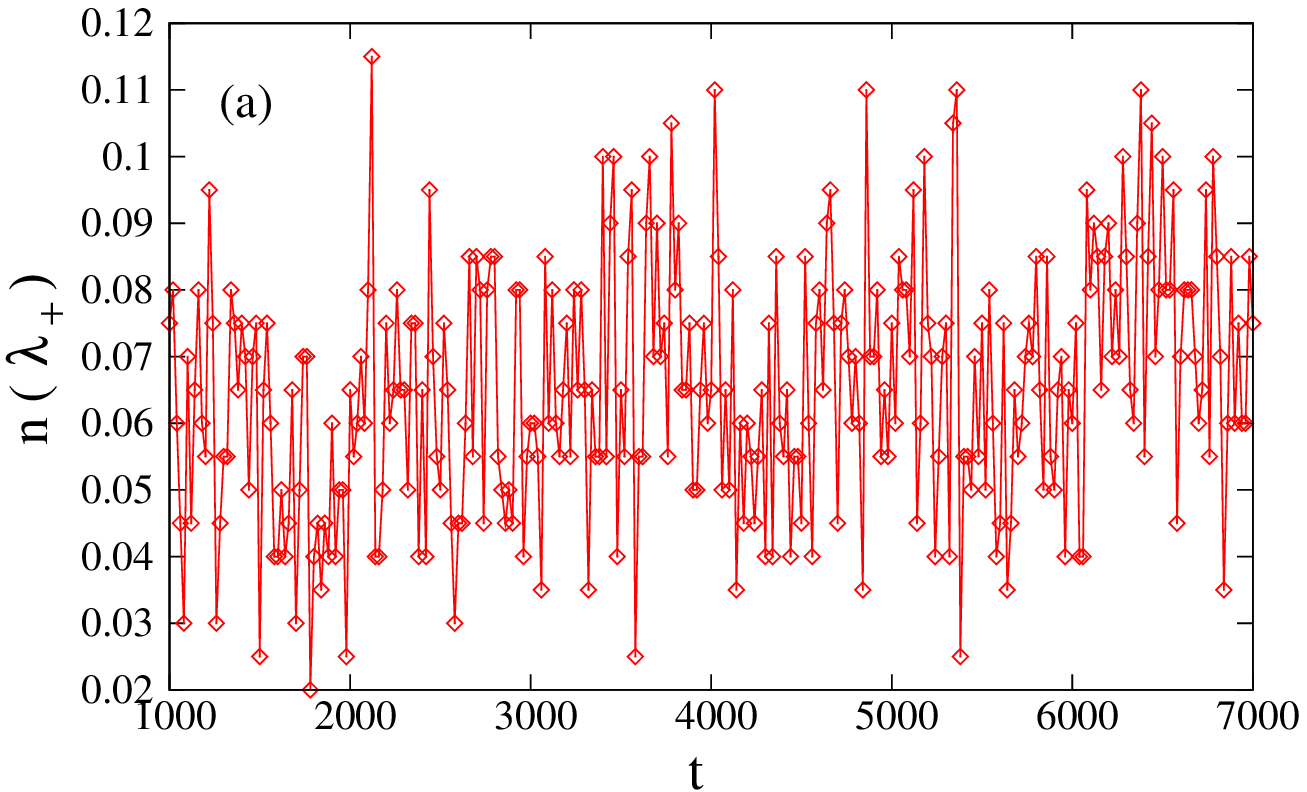}&
\includegraphics[scale=0.6]{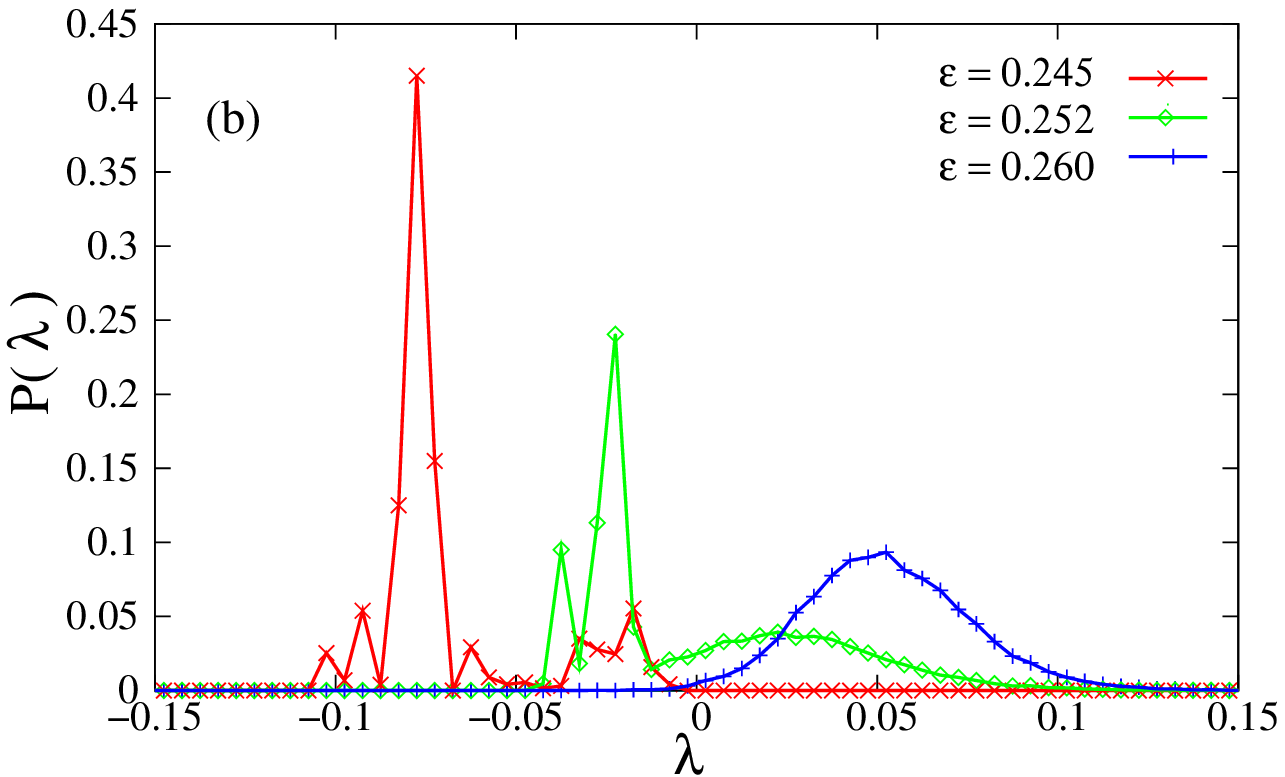}\\
\end{tabular}
\end{center}
\caption{(Color online) shows (a) the fraction of positive finite-time Lyapunov exponents (time=20) as  a function of time $t$ obtained at $\Omega=0.065, \epsilon=0.254$, and (b) the distribution of the largest finite-time Lyapunov exponent (time$=100$) plotted for $\Omega=0.065$ and $\epsilon=0.245, 0.252, 0.260$. \label{ftle}}
\end{figure}

For the coupled map lattice of eq. \ref{evol}, which is an N-dimensional map of the form $\mathbf{x}\rightarrow\mathbf{f(x)}$, let the Jacobian evaluated at the initial condition $\mathbf{x}_0$ be denoted as $\mathbf{Df(x_0)}$. 
 The time-$n$ Lyapunov exponents (FTLE) for this system, are obtained by diagonalizing the Jacobian  $\mathbf{Df^n(x_0)}$ of the n-times iterated map $\mathbf{f^n}$,  the $k^{th}$ time-$n$ Lyapunov exponent being defined as 
\beq
\lambda_k(\mathbf{x_o},n)=\frac{1}{n}\log \Lambda_k
\label{ftle_eq}
\eeq
where, $\Lambda_k$ is the  $k^{th}$ eigenvalue of the time-$n$ Jacobian $\mathbf{Df^n(x_0)}$. 

We find that the  finite-time Lyapunov exponents obtained for our system exhibit fluctuations about zero in the vicinity of the infection line. This can be seen in Fig. \ref{ftle}(a), where the fraction of positive time-20 Lyapunov exponents obtained at $\Omega=0.065, \epsilon=0.254$, is plotted as a function of time. The fraction of positive exponents is seen to vary with time, indicating that  a fraction of the Lyapunov exponents fluctuate about zero, as the dynamics evolves. Hence, the trajectory of the coupled map visits periodic orbits with different number of unstable directions, in the vicinity of the infection line. 

This is further illustrated in Fig. \ref{ftle}(b), which shows the distribution of the largest time-100 Lyapunov exponent, obtained at $\Omega=0.065$  for various values of $\epsilon$. The distribution has been calculated by collecting data for  a lattice of size $N=200$ over $10000$ time steps and by averaging over $200$ initial conditions. We find that the distribution of the largest time-100 Lyapunov exponent is constrained to the negative values for parameters lying in the non-spreading regime (eg. $\epsilon=0.245$ in Fig. \ref{ftle}(b)). For parameters close to the infection line (viz. $\epsilon=0.252$), the distribution shifts towards the positive side, until it lies entirely in the positive side of the axis for parameters above the infection line (eg. $\epsilon=0.260$). Hence, the distribution of Lyapunov exponents obtained at $\epsilon=0.252$ illustrates further that the largest finite-time Lyapunov exponent vacillates around zero near the infection line.

 Therefore, we infer from Fig. \ref{ftle}(a) and (b), that the finite-time Lyapunov exponents fluctuate about zero, near the infection line, thereby confirming the presence of unstable dimension variability at the infection line. Hence, we see that the spreading nature of the bursts above the infection line can be attributed to the  attractor-widening crisis as the well as the presence of unstable dimension variability at the infection line. 
 
 \section{Discussion}

To conclude,
the spatiotemporal intermittency of the directed percolation class and
the spatial intermittency of the non-directed percolation
class, seen along the bifurcation boundaries of spatiotemporally fixed point solutions,
are special cases of the spreading and non-spreading regimes seen off the
bifurcation boundaries. The two regimes are separated by the infection
line which intersects with the bifurcation boundary of the fixed point 
solutions at the point where the cross-over between the directed
percolation and non-directed percolation
behavior takes place. Thus the behavior seen in coupled sine circle map
 lattice is organized around the locations of the bifurcation boundaries of
the fixed point solutions, and the infection line. The existence of two
distinct universality
classes,
in the phase diagram of the sine circle map lattice is 
reflected in  the  transition of the equivalent
cellular automaton from the probabilistic phase to the deterministic
phase and the concomitant suppression of the spreading or
infectious modes. The dynamic origins of this transition lie in an
attractor-widening crisis. We also find evidence for the existence of
unstable dimension variability in the neighborhood of the infection line.
Thus, in this system, there is a 
direct connection between a dynamical phenomenon viz. a crisis
in an extended system and the statistical properties of the extended
system viz. the exponents and universality classes. Similar directed
percolation to non-directed percolation  transitions have been seen in
other coupled map lattices, as well as in pair contact processes, solid on
solid
models  and models of non-equilibrium wetting \cite{Odor}.
Our results may  have useful pointers for the analysis of other
systems, and thus contribute to the on-going
debate on the identification of the universality classes of
spatiotemporal systems.

Acknowledgment: NG thanks DST, India for partial support under the project SR/S2/HEP/10/2003.


\begin{thebibliography}{99}
\bibitem{sti_papers} 
F. Daviaud, M. Bonetti and  M. Dubois,  Phys. Rev. A. {\bf 42}, 3388 (1990); P. W. Colovas and C. D. Andereck, Phys. Rev. E. {\bf 55}, 2736 (1997);M. Das, B. Chakrabarti, C. Dasgupta, S. Ramaswamy and A.K. Sood, Phys. Rev. E. {\bf 71}, 021707 (2005);C. Pirat,  A. Naso, J.L Meunier, P. Maïssa and C. Mathis, Phys. Rev. Lett. {\bf 94}, 134502 (2005).

\bibitem{pomeau}  Y. Pomeau, Physica D. {\bf 23}, 3 (1986).

 \bibitem{chate} 
H. Chat\'e and P. Manneville, Physica D {\bf 32}, 409 (1988).
\bibitem{grassberger}
P. Grassberger and T. Schreiber, Physica D,{\bf 50},177(1991).
\bibitem{bohr}
T. Bohr, M. van Hecke, R. Mikkelsen, M. Ipsen Phys. Rev. Lett. 86 5482 (2001).
\bibitem{rolf} J. Rolf, T. Bohr, and M. H. Jensen, Phys. Rev. E {\bf 57}, R2503 (1998).
\bibitem{rupp} P. Rupp, R. Richter and I. Rehberg, Phys. Rev. E. {\bf 67}, 036209 (2003).
\bibitem{takeuchi} K.A. Takeuchi, M. Kuroda, H. Chat\'e, and M. Sano, Phys. Rev. Lett {\bf 99}, 234503 (2007).
\bibitem{janaki}
T.M. Janaki, S. Sinha, and N. Gupte, Phy. Rev. E {\bf 67}, 056218 (2003).
\bibitem{zjngpre72}
Z. Jabeen and N. Gupte, Phys. Rev. E {\bf 72}, 016202(2005).
  \bibitem{zjngpre74}
Z. Jabeen and N. Gupte, Phys. Rev. E {\bf 74}, 016210(2006).


 \bibitem{gauri2}
  G. Pradhan, N. Chatterjee, and N. Gupte, Phys. Rev. E 65, 046227 (2002).

 \bibitem{ashutosh} 
A. Sharma and N. Gupte, Phys. Rev. E 66, 036210 (2002). 


\bibitem{domanykinzel} 
 E. Domany and W. Kinzel, Phys. Rev. Lett.{\bf 53},311(1984).

\bibitem{notep2}
The probability  $p_2=P(1|010)$ is also seen to be equal to zero in both the regimes, though the reasons are different. In the spreading regime, we find that the neighboring laminar states  suppress the central burst state, and hence we obtain a zero probability for $p_2$. In the case of the non-spreading regime, a single burst site with two laminar neighbors is never observed.

\bibitem{bagnolica} A similar study has been carried out in 
F. Bagnoli {\it et al},  Phys. Rev. E. {\bf 63}, 046116 (2001),  in which the probability rules were defined differently.
  

\bibitem{yorke1} C. Grebogi, E. Ott, and J.A.Yorke, Phys. Rev. Lett. {\bf 48}, 1507 (1982).
\bibitem{greb} C. Grebogi, E. Ott, F. Romeiras, and J.A.Yorke, Phys. Rev. A {\bf 36}, 5365 (1987).
\bibitem{kost} E.J.Kostelich, I. Kan, C. Grebogi, E.Ott, and J.A.Yorke, Physica D {\bf 109}, 81 (1997).
\bibitem{lai}
Y.C. Lai, C. Grebogi and J. Kurths,Phys. Rev. E {\bf 59}, 2907 (1999);Y.C.Lai, D. Lerner, K. Williams, and C. Grebogi, Phys. Rev. E {\bf 60}, 5445 (1999).


\bibitem{Odor}
G. Odor, Rev. Mod. Phys. {\bf 76}, 663, (2004). 

\end{thebibliography}
\end{document}